\documentclass{article}
\usepackage[T1]{fontenc}
\usepackage[latin2]{inputenc}
\usepackage{latexsym}
\usepackage{amsfonts}
\usepackage{amsmath}
\usepackage{verbatim}
\usepackage{hyperref}
\newcommand{\de}{\mathrm{d}}

\title{The continuum approach to the BF vacuum: the U(1) case}
\author{Patryk Drobi\'nski$^1$ \and Jerzy Lewandowski$^1$}
\date{\small{$^1$Faculty of Physics, University of Warsaw, ul. Pasteura 5, 02-093 Warszawa, Poland}\\[2ex]
\today}
\begin{document}
\maketitle
\begin{abstract}
A quantum representation of holonomies and exponentiated fluxes of a $U(1)$ gauge theory that  contains the Pullin-Dittrich-Geiller (DG) vacuum 
is presented and discussed. Our quantization is performed manifestly in a continuum theory, without any discretization. The discretness
emerges on the quantum level as a property of the spectrum of the quantum holonomy operators. The new type of a cylindrical consistency
present in the DG approach, now follows easily and naturally. A generalization to the non--Abelian case seems possible.  
\end{abstract}
\tableofcontents
\section{Introduction}
Loop quantum gravity (LQG) \cite{LQG} is a background independent quantum gravity theory, based on  the methods of canonical quantization. The kinematical Hilbert space of quantum states is built around geometry excitations on a vacuum which corresponds to a completely vanishing spatial quantum metric tensor \cite{AL}. It has become a staple of loop quantum gravity with the powerful LOST theorem \cite{LOST} stating that this vacuum state is unique, given certain natural assumptions. However, an outstanding problem in this approach is distinguishing a quantum flat spacetime state.  Hence, other possible vacuums have been  looked into.  Pullin \cite{Pullin} considered a state supported at an everywhere vanishing Ashtekar connection. He pointed out that in spite of the simple form the state defines a  non--trivial spatial metric tensor (see also \cite{BLM}). A similar idea was brought up recently and developed further by   Dittrich and Geiller. They introduced a corresponding quantum representation of holonomies and fluxes in 2+1 dimensions \cite{dittrich1} and later extended it to 3+1 dimensions \cite{dittrich2,dittrich3}. The gauge group is assumed to be a compact semi-simple Lie group or a finite group. Since their vacuum corresponds to a vanishing connection, it is  referred to as the BF vacuum.  It was derived by using the framework of the discretized  BF theory and introducing a new cylindrical consistency condition between different discretizations, a requirement ensuring the existence of a continuum limit. New quantum states for LQG were also proposed 
by Sahlmann and one of the authors of the current paper  \cite{lewandowski}.  The main emphasis was on the scalar field, however a two line derivation of the DG vacuum was also mentioned in that work. In the current paper we extend that idea. Our framework is defined directly for a continuum theory. We derive from it the DG approach without any discretization. Our results are valid for the gauge group $U(1)$ or any other Abelian Lie group. However,  tools necessary for a generalization to a non--Abelian group are known and available. 
  
\section{$U(1)$ connections and conjugate momenta}
On an $n$-manifold $\Sigma$ we consider the fields of canonical abelian gauge theory, namely a vector field 
\begin{equation}
A\ =\ A_adx^a
\end{equation} 
and  conjugate vector density
\begin{equation}
E \ =\ \frac{1}{(n-1)!}E^a \epsilon_{aa_1...a_{n-1}}dx^1\wedge...\wedge dx^{n-1} 
\end{equation}
that satisfy the kinematical Poisson bracket 
\begin{equation}
\{A_a(x),E^b(y)\}\ =\ \delta_a^b\delta(x,y) .
\end{equation} 
They are also subject to gauge transformations 
\begin{equation}
(A,\ E)\ \mapsto\ (A + d\lambda,\ E)
\end{equation} 
The generator of the gauge transformations is  the Gauss constraint function
\begin{equation}
G\ :=\ \partial_aE^a .
\end{equation} 
The physical fields satisfy the constraint
\begin{equation}
G\ =\ 0
\end{equation} 
that equivalently reads
\begin{equation}\label{cnstr}
dE\ =\ 0\ .
\end{equation} 
In the presence of the constraint there are two equivalent procedures: either the Poisson bracket should be replaced 
by the suitable Dirac bracket or we should use manifestly gauge invariant observables, that is functions  of the fields such that
\begin{equation}
{\cal O}(A+d\lambda,E)\ =\ {\cal O}(A,E) 
\end{equation} 
defined on the space of unconstrained fields and after calculating the Poisson bracket
restrict them to the constrained data. We chose the latter option.

Obvious examples of gauge invariant functions of $A$ are given by components of the curvature,
\begin{equation}
F_{ab}\ =\ \partial_aA_b - \partial_bA_a .
\end{equation}
 
Another example are the holonomy functions. Every loop $\gamma$  in $\Sigma$ defines a 
holonomy observable
\begin{equation}
h_\gamma(A) \ =\  e^{-i\int_\gamma A} \label{holo} 
\end{equation}
that is gauge invariant
\begin{equation}
h_\gamma(A+d\lambda) \ =\  h_\gamma(A) . 
\end{equation}

Since $E$ is gauge invariant, so is every function of $E$.  However, in this paper we will use
very special ones.  Every 1-form 
\begin{equation}
\omega\ =\ \omega_b dx^b
\end{equation}
defines an observable, a function of $E$,     
\begin{equation}
e^{-i \int \omega\wedge E} .\label{flux}
\end{equation}  
Since we consider this function on the space of solutions to the constraint (\ref{cnstr}), 
we have
\begin{equation}
e^{-i \int (\omega+d\lambda)\wedge E} \ = e^{-i \int \omega\wedge E} ,
\end{equation} 
meaning that the smearing 1-form $\omega$ is determined up to the transformations
\begin{equation}\label{gaugeomega}
\omega\ \mapsto\ \omega+d\lambda 
\end{equation} 
with an arbitrary function $\lambda$ defined on $\Sigma$. 
 
The exponentiated momentum functions set an algebra with a simple multiplication law
\begin{equation}
e^{-i \omega_1\wedge E} e^{-i \omega_2\wedge E}\ = e^{-i \int (\omega_1+\omega_2)\wedge E} .
\end{equation}  
 
The smooth smearing $1$-forms $\omega$ used above can be generalized to  distributions
supported on co-dimension $1$ complexes. Let $\Delta$ be a  triangulation of $\Sigma$.
Consider the $(n-1)$-complex defined by the $n-1$ dimensional faces of $\Delta$ endowed
with orientation. We color each face $f$ of $\Delta$ by a constant,
\begin{equation}
f\mapsto \omega_f \in \mathbb{R} .
\end{equation} 
The corresponding distributional $1$-form $\omega$ is defined by an integral with an arbitrary field $E$, namely
\begin{equation} \label{distromega} 
\int_\Sigma \omega\wedge E\ :=\ \sum_f \omega_f\int_f E , 
\end{equation}
Finally, we consider the exponential function
\begin{equation}\label{1}
E\ \mapsto\   
e^{-i\sum_f \omega_f\int_f E}
\end{equation} 
 
Two different  partitions $\Delta$ and $\Delta'$  may give rise to the same distributional 1-form $\omega$ (\ref{distromega}), and in the consequence the same observable (\ref{flux}).  An identical function 
\begin{equation}
E\ \mapsto\  \sum_f \omega_f\int_f E
\end{equation} 
can be obtained from a refinement $\Delta'$ of the partition $\Delta$. To every  face $f'$
of $\Delta'$ obtained by dividing a face $f$ of $\Delta$ we assign
\begin{equation} \label{refin1}
\omega'_{f'}\ :=\ \omega_f .
\end{equation}  
To a face $f'$ of $\Delta'$ that is not obtained in that way, and therefore does not overlap
any of the unrefined faces, we assign
\begin{equation} 
\omega'_{f'}\ :=\ 0.  \label{refin2}
\end{equation}  
Indeed, then
\begin{equation}
 \sum_f \omega_f\int_f E\ =\ \sum_{f'} \omega'_{f'}\int_{f'} E
\end{equation} 
for every field $E$.
Another possibility is a partition $\Delta'$ obtained  from $\Delta$ by flipping the orientation of a face $f_1$ and calling it $f'_1$.  
We assign to it consistently
\begin{equation}  \label{flip}
\omega'_{f'_1}\ :=\ -\omega_{f_1}.
\end{equation}  
 Again, for every field $E$
\begin{equation}\label{2}
 \sum_f \omega_f\int_f E\ =\ \sum_{f'} \omega'_{f'}\int_f E
\end{equation} 

Notice that a general  "gauge" transformation (\ref{gaugeomega}) of $\omega$   (at this point $\omega$ is not a connection, therefore is not subject
to the gauge transformations of that theory) changes the support of the distribution $\omega$ unless, inside every simplicial cell $\triangle$
of the triangulation $\Delta$, 
\begin{equation}  
d\lambda_\triangle\ =\ 0 .
\end{equation}  
That restricts considerably the ambiguity (\ref{gaugeomega}), however, not completely.
Let $\lambda$ be constant on  every cell $\triangle$ of $\Delta$, but possibly change value  whenever we pass from one cell $\triangle_1$ to another $\triangle_2$ across a face $f$ they share. Then, the corresponding 
transformation (\ref{gaugeomega})  amounts to
\begin{equation}   \label{gauge}
\omega'_f\ =\ \omega_f + \delta\lambda_f ,
\end{equation}
where $\delta\lambda_f$ is the jump of $\lambda$ (the orientation of $f$ defines the sign of the jump).      

As in the case of smooth $\lambda$, for every field $E$ that satisfies the constraint (\ref{cnstr})
\begin{equation}\label{3}
 \sum_f \omega_f\int_f E\ =\ \sum_{f} (\omega_{f}+\delta\lambda_f)\int_f E ,
\end{equation} 
 owing to 
 \begin{equation}
 \int_{\partial\triangle}E\ =\ 0 .
 \end{equation}
 
 Incidentally, for each trangulation $\Delta$,  the distributional ambiguity transformations (\ref{gauge}) could be restricted 
 by suitable  gauge fixing. The standard procedure is to use any graph dual to $\Delta$ (topologically, the dual graph is unique,
 however here we consider embedded graphs, and there are infinitely many embeddings) and choose a maximal tree. We could
 use the transformations (\ref{gauge}) to fix 
 \begin{equation}   
\omega'_f\ =\ 0,
\end{equation}
 for every face $f$ intersected by the graph $\Gamma$ and impose a gauge condition for the nonzero $\omega$'s. 
 However, this will not be necessary.        
 
 Now, what about the product of two exponentiated fluxes, namely
 \begin{equation}
e^{-i \omega^1\wedge E} e^{-i \omega^2\wedge E}\ = e^{-i \int (\omega^1+\omega^2)\wedge E} .
\end{equation}  
If $\omega^1$ and $\omega^2$ are defined by two different colorings of the faces of a single triangulation $\Delta$,
then their sum $\omega^1+\omega^2$ corresponds to a coloring 
\begin{equation} \label{sum}  
\omega^3_f\ =\ \omega^1_f\ +\ \omega^2_f.
\end{equation}
The problem is less trivial, if $\omega^1$ is represented by a coloring  of a triangulation $\Delta^1$, while
$\omega^2$ is represented by a coloring of a different triangulation $\Delta^2$. 
It is solved provided there exists a common refinement $\Delta^3$ of the complexes $\Delta^1$ and $\Delta^2$.
Then, we first find the colorings of $\Delta^3$ corresponding  to $\omega^1$ and $\omega^2$ and next apply (\ref{sum}).
In the category of smooth manifolds such a common refinement may not exist. It exists in the semi--analytic 
category  (see \cite{LOST}).
 
\section{The quantum momentum representation}
 The quantum momentum  representation amounts to promoting each functional  $e^{-i\int \omega\wedge E}$
to a state   $|a\rangle$, where
\begin{equation}
a\ :=\ \hbar \omega .
\end{equation}
That is, the state is a function
\begin{equation}
|a\rangle:\ \ E\ \mapsto\  e^{-\frac{i}{\hbar}\int a\wedge E} .
\end{equation}
invoking  (\ref{gaugeomega}) we should remember that a given state $|a\rangle$ determines $a$ up to the transformations
\begin{equation}\label{gaugea}
|a\rangle \ =\ |a+d\lambda\rangle .
\end{equation}

Quantum operators
\begin{equation} \label{hatA}
\hat{A}_a(x)= {i\hbar}\frac{\delta}{\delta E^a(x)}\ , \ \ \  \ \ \ \ \ \ \  \hat{E}^a(x)\ =\ E^a(x),
\end{equation} 
will not be defined in our representation themselves, however we use them as auxiliary operations 
to define the proper quantum operators
\begin{align}\label{hol}
\widehat{h_\gamma({A})}|a\rangle\ :=\   h_\gamma(\hat{A})|a\rangle\ =\  h_\gamma(a) |a\rangle\\
\widehat{e^{-i \int \omega\wedge {E}}}|a\rangle\ :=\  {e^{- i \int \omega\wedge \hat{E}}}|a\rangle
 \ =\ |a+\hbar \omega\rangle.\label{flux}
\end{align}
In conclusion, the label $a$ defining a quantum state gets the interpretation of an eigenvalue of 
the quantum connection, or more precisely, for every loop $\gamma$ the exponentiated integral
$$ e^{-i\int_\gamma a }$$
is the eigenvalue  of the corresponding quantum holonomy operator $h_\gamma(\hat{A})$. 
This is consistent with the ambiguity (\ref{gaugea}), so $a$ defines a gauge equivalence class of connections. 
In this representation $a$ is a smooth 1-form, hence the corresponding connections are regular classical ones.    

Incidentally,   also the field strength
\begin{equation}
 F\ =\ dA
\end{equation}
quantum operator  $\hat{F}$ is well defined in this representation
\begin{equation}
\hat F_{bc}|a\rangle\ =\   \hbar(\partial_ba_c-\partial_ca_b)|a\rangle . \end{equation}
The Hilbert product may be defined as follows
\begin{equation}\label{product}
\langle a|a'\rangle=\left\{\begin{array}{rl}
1&\text{if }a=a'\ \ \ \ \ {\rm mod}\ \ \ d\lambda,\\
0&\text{if }a\neq a'\ \ \ \ \ {\rm mod}\ \ \ d\lambda.
\end{array}\right.
\end{equation}

\section{The flux representation}
The flux representation is a quantization of the exponentiated flux observables (\ref{1}) and all the holonomy observables.
The idea is to use the formulae of the exponentiated momenta representation, and make sure they extend to the
fluxes.  To start with, the exponentiated fluxes (\ref{1}) are used for the construction of the quantum state. A function 
\begin{equation}\label{a}
 E\ \mapsto\  e^{-\frac{i}{\hbar}\sum_f a_f\int_f E}
\end{equation} 
where $f$ ranges the set of faces of a given oriented triangulation $\Delta$ and $f\mapsto a_f$ is a 
coloring, is now promoted to a quantum state $|a\rangle$.
It follows from the discussion in the previous section that given a state $|a\rangle$,  the colored oriented triangulation
$\Delta, (f\mapsto a_f)$ is determined modulo arbitrary compositions of the following steps: 
\begin{itemize}
\item a refinement (\ref{refin1}, \ref{refin2}), 
\item flipping the orientation (\ref{flip}), 
\item gauge transformations  (\ref{gauge})  
\end{itemize}
characterized above with the substitution 
$$\omega\ \mapsto \ \frac{a}{\hbar}, \ \ \ \ \ \ \  {\rm etc.}$$ 

The Hilbert product is still defined by (\ref{product}).  

We define the action of the quantum holonomy operators on  the flux states again by  the formulae (\ref{hatA},\ref{hol}).  
Remarkably,  the holonomy map
\begin{equation}\label{hol}
\gamma\ \mapsto\ e^{-i\int_\gamma a}\ =:\ h_\gamma(a)
\end{equation}
is still well defined for every  generalized, distributional 1-form $a$ defined by (\ref{distromega}) and
$$ a\ :=\ \hbar{\omega}, \ \ \ \ \ \ \ \ \ \ \ \ a_f\ :=\ \hbar \omega_f .$$

Indeed, for each loop $\gamma$ that has a generic intersection with the triangulation $\Delta$,  
\begin{equation} \label{hgamma} 
h_\gamma(a) \ =\ e^{-i\sum_f n_f a_f}
\end{equation}
where for every face $f$,  $n_f$ is the number of times $\gamma$ intersects the face $f$ 
in the direction consistent with the orientation of $f$ minus the number of times $\gamma$ intersects the face $f$
in the direction opposite to the orientation of $f$.  The intersection $\gamma\cap \Delta$ is called generic if it is a finite
set, such that every intersection point is contained in exactly one face $f$ of $\Delta$, and the loop $\gamma$ passes 
transversally through $f$ at that point.  Degenerate  intersections on the other hand,  require a regularization.  Given 
a segment of $\gamma$ that has a degenerate intersection with $\Delta$, we replace it by a small deformation that 
intersects $\Delta$ generically. Since a deformation is not unique, we take the arithmetic mean of the resulting set of distinct values 
$\sum _f n'_f a_f$ in (\ref{hgamma}). Each total sum is gauge invariant, and invariant with respect to the refinements of
$\Delta$. Therefore, neither gauge transformations nor refinements change that set of values. In that way we find that 
if $\gamma$ passes through  an edge of a face of $\Delta$, then generically (in this class of degenerate intersections) 
that intersection contributes  $\pm\frac{1}{2}a_f$. In the case of $\gamma$ passing through a vertex of a face $f$, 
depending on other faces $f'$ meeting at that vertex and their charges $a_{f'}$, the contribution of the face $f$ may 
be any fraction of $a_f$.    
          
That generalized, distributional connection $a$ defines a map (\ref{hol}) from the space of loops into the group $U(1)$, 
that has all the algebraic properties of the holonomy map of a regular connection, that is
\begin{equation}  
h_{\gamma\circ\gamma'}\ =\ h_{\gamma}h_{\gamma'}, \ \ \ \ \ \ \ \ \ h_{\gamma^{-1}}\ =\  (h_{\gamma})^{-1}  
\end{equation}
where $\gamma^{-1}$ is obtained from $\gamma$ by flipping the orientation, and the last "$-1$" is in the sense
of $U(1)$.  An example of a generalized in this sense connection supported on 2--surfaces on a 3--dimensional manifold 
was given in \cite{lewandowski}.                

We now turn to the discussion of the properties of the generalized connections defined above. The only way for the loop to define a non-trivial  
holonomy  is  to encircle or intersect one or more edges of the complex.  In this sense, the curvature is supported at the edges. 
For example a loop $\gamma_1$ going around a single edge defines a holonomy
\begin{equation}  
h_{\gamma_1}(a) \ =\ e^{-i\sum_i \pm a_{f_i}}\label{holonomia}
\end{equation}
where $f_i$ ranges over the set of the faces containing the edge. On the other hand, a loop $\gamma_2$ 
which is all contained in a single cell defines
\begin{equation}    
h_{\gamma_2}(a) \ =\ e^{i0}\ =\ 1 .
\end{equation}
Also a loop $\gamma_3$ that goes from one cell into another one across a shared face $f_0$
and comes back intersecting $f_0$ another time has the holonomy
\begin{equation}  
h_{\gamma_3}(a) \ =\ e^{\pm i\sum a_{f} - a_f}\ =\ 1.
\end{equation}

Finally, for the vacuum state $|0\rangle$, for every loop $\gamma$
\begin{equation}\label{0}  
h_{\gamma}(0)\ =\ 1 .
\end{equation}

Surprisingly, though, there are more states of that property. Consider a state $|a\rangle$, defined by a coloring
\begin{equation}\label{vacua}
f\ \mapsto\ 2\pi m_f , \ \ \ \ \ \ \ \ m_f\in \mathbb{Z} .
\end{equation}
The choice of the integers $m_f$ is arbitrary, so there are many such states.  For each of them (\ref{0}) holds for every
generic curve $\gamma$.   Non-generic curves can have holonomies with fractions of $a_f$, though. However, even
those curves give trivial holonomy if the numbers $m_f$ have sufficiently many divisors. In conclusion, there exist states $|a\rangle$ (\ref{vacua}) with $m_f\not=0$ such that  the action of every quantum holonomy operator is trivial.

An operator  $e^{-i\int\omega\wedge \hat{E}}$  is defined again by (\ref{hatA}). It  maps the function (\ref{a})
into a new function
\begin{equation}
 E\ \mapsto\  e^{-\frac{i}{\hbar}(a+\hbar\omega)\wedge E}
\end{equation} 
where the distributions $a$ and $\hbar\omega$ can be naturally added. As it was explained in the previous section,
the sum $a+\hbar\omega$ can also be represented by an ordered  triangulation $\Delta$ and coloring of the faces
by real numbers. Therefore 
\begin{equation}  
e^{-i\int\omega\wedge \hat{E}}|a\rangle\ =\ |a+\hbar\omega\rangle
\end{equation}
maps quantum flux states into quantum  flux states. 

In conclusion, the  flux representation is a quantization of the observables \eqref{holo},\eqref{flux} defined on the space 
of $U(1)$ connections and canonically conjugate momenta. From the beginning one deals with the full set of the local degrees of freedom.  
What emerges as eigenvalues of the quantum holonomy operators are generalized connections  (modulo the gauge transformations) characterized in this section.     
They have a discretized character, however they emerge as elements of quantum theory in continuum.

\section{The BF vacuum}
In this section we briefly perform the construction of Dittrich and Geiller, however we replace the non-abelian gauge group by the group $U(1)$.  
We  stick to $3$ dimensional space $\Sigma$, as it is done in the original DG framework, although generalization to any number of dimensions 
is straightforward at this point.
\subsection{The discretization and quantum states}
Again,  the starting point is a 3--dimensional triangulation $\Delta$ of $\Sigma$.  But in this approach, already at the classical level,  we restrict ourselves to the 
discretized distributional connections  flat inside  each 3--simplex of $\Delta$.  These are the generalized connections introduced in the previous section. Let us call them 
  compatible with $\Delta$.  To characterize uniquely  each connection  it is sufficient to fix  a finite set of $N$ curves,  each one intersecting transversally one face $f$ of $\Delta$ and consider holonomies along them. We choose those curves, say $e_1,...,e_N$, such that they form an embedded  graph $\Gamma$ dual to $\Delta$ and the edges are oriented consistently with the faces.  Every generalized connection $A$ 
  compatible with $\Delta$  is characterized by the sequence of  holonomies
 $$ (g_1,...,g_N)\ =\ (h_{e_1}(A), ..., h_{e_N}(A)) . $$
 
The Hilbert space associated with this triangulation is a subspace of the space of complex functions $\psi$ of holonomies along the edges of $\Gamma$:
\begin{equation}
\psi:G^N\mapsto\mathbb{C}
\end{equation}
It is spanned (and later completed) by functions labelled by sequences $(\alpha_1,\ldots,\alpha_N)$ of $U(1)$ elements,  namely
\begin{equation}\label{psi}
\psi_{\{\alpha_i\}}(g_1,\ldots,g_N)=\prod_{i=1}^N\delta_{g_i,\alpha_i}
\end{equation}
with $\delta_{\alpha,\beta}=1$ when $\alpha=\beta$ and 0 otherwise. 

\subsection{Refinement of the Hilbert space}
The BF representation carries a natural notion of refinement. Consider a triangulation $\Delta$  and a refined triangulation $\Delta'$, which can be obtained by subdividing some of the 2--simplices of $\Delta$ into a number of smaller ones. The corresponding dual graphs are denoted by 
$\Gamma$ and $\Gamma'$, respectively.  A basis state $\psi_{\{\alpha_i\}}$ associated to  $\Delta$
is identified with  a unique state $\psi'_{\{\alpha_i'\}}$ associated to $\Delta'$ in the following way. For every path $\gamma'$ in the graph $\Gamma'$ dual to $\Delta'$ (not necessarily a closed loop) we define its projection 
$P_\Gamma(\gamma')$ as the unique path in $\Gamma$ which crosses the faces of $\Delta$ in exactly the same order as the path $\gamma'$. 
Next, notice, that the sequence $(\alpha_1,...,\alpha_n)$ defines uniquely a connection $A$ compatible with $\Delta$. 
For $\Delta'$ we find a unique set ${\alpha'_1,...,\alpha'_{N'}}$ such that the  corresponding connection $A'$ compatible with $\Delta'$ 
satisfies the constraint that the holonomies along the path and its projection are equal:
\begin{equation}\label{comp}
h_{\gamma'}(A') =h_{P_\Gamma(\gamma')}(A).
\end{equation}
 The proofs that this refinement procedure results in a unique and well--defined state can be found in \cite{dittrich3}. As a matter of fact, the condition (\ref{comp}) means that $A'$ simply equals $A$.
\subsection{Holonomy and flux operators}
Given a triangulation $\Delta$ and the space spanned by the quantum states $\psi_{\{\alpha_i\}}$ labelled by the sequences $(\alpha_1,...,\alpha_N)\in U(1)^N$, we define the quantum operator ${h}_{e_i}(\hat{A})$ of the holonomy along an edge $e_i$  to act as follows
\begin{equation}
{h}_{e_i}(\hat{A})\psi(\{g_j\})\ :=\ g_i\psi(\{g_j\}).
\end{equation}
The basic observables for the $E$ field    are the  fluxes through the faces $f_i$ of $\Delta$ corresponding to the edges $e_i$ of $\Gamma$, defined as
\begin{equation} 
X_i\ :=\int_{f_i}E,\label{Xe1} .
\end{equation}
The operator corresponding to an observable $X_i$ would act on states as a derivative over the argument corresponding to the edge $e$:
\begin{equation}
\hat{X}_i\psi(g_1,\ldots,g_i,\ldots,g_N)=\frac{\de}{\de t}\psi(g_1,\ldots,g_ie^{it},\ldots,g_N)|_{t=0}.
\end{equation}
 However, such operators would not be well--defined on the basis states, so they need to be exponentiated instead.
\subsection{Map between the representations}\label{sec5}
There is a natural map that carries the states of the exponentiated flux representation into the DG states. 
Given a state $|a\rangle$ and the corresponding generalized connection $a$, for every triangulation $\Delta$ and the dual 
graph $\Gamma$ used in the BF vacuum formalism, define  
\begin{equation}\label{themap}
|a\rangle\ \mapsto\ \psi_{\{h_{e_i}(a)\}} 
\end{equation}
where  $e_1,...e_N$ are edges of $\Gamma$ and $h_{e_i}(a)$ is the classical holonomy along $e_i$ with respect
to the generalized connection $a$. If $\Delta'$ is a refinement of $\Delta$, the corresponding states 
 $\psi_{\{h_{e_i}(a)\}}$   and $\psi_{\{h_{e'_{i'}}(a)\}}$  are consistent in the sense of the previous section. 

The  map (\ref{themap}) however has a non-trivial kernel. Indeed, two different states  $|a\rangle$ and $|a'\rangle$
are carried into a single state whenever  corresponding colorings of a common triangulation  are related to each other
by adding multiples of $2\pi$,
\begin{equation}\label{ambi}
a'_f\ =\ a_f + m_f2\pi, \ \ \ \ \ \ \ \ \ \ \ \ m_f\in\mathbb{Z} .
\end{equation}
The exponentiated quantum flux operators are intertwined by the map (\ref{ambi}). 
Intertwining of the quantum holonomy operators is not so simple and 
in fact does not take place. On one hand, given a state $|a\rangle$  defined using a triangulation 
$\Delta$ and a generic loop,  the transformation (\ref{ambi}) does not change the eigenvalue of the corresponding quantum 
holonomy operator because every $n_f$ in (\ref{hgamma}) is an integer. Therefore for that $|a\rangle$ 
and a generic with respect to it loop $\gamma$, the quantum holonomy operator  $\hat{h}_\gamma$
passes through the map (\ref{themap}). However, $n_f$ may be half integer in the case of a loop that
intersects an edge of $\Delta$. Then, our quantum holonomy operator can tell between  $|a\rangle$ and 
some of $|a'\rangle$ obtained by (\ref{ambi}).  
 
\section{Summary}
In summary, we have discussed in this paper a quantization valid for the canonical fields of  $U(1)$ gauge theory
on a space of arbitrary dimension. The classical observables we directly turn into quantum operators are holonomies
(\ref{holo}) along arbitrary loops in the space, and exponentiated fluxes (\ref{1}) of the electric field.  Importantly, our framework 
is formulated manifestly in the continuum.  There is no discretization involved.  Quantum states are functions of the exponentiated fluxes
defined in the space of all the electric fields subject to the Gauss constraint. The eigenvalues of the quantum holonomy 
operators set holonomies of some generalized   connections supported on the co-dimension $1$ faces of triangulations
of the underlying 3-dimensional manifold. 
Their structure is  partially  discrete. Each of the quantum generalized connections can be characterized by
a triangulation and a coloring of its faces.  Our quantum representation is compared to the representation proposed 
by Dittrich and Geiller. The latter one relies on a discretization performed  as the starting point already at the classical level. 
There, the distributional connections   are put in by hand as discretized classical degrees of freedom. 
We construct a natural  map between the two  representations. The correspondence however is not 1-1. The map we construct 
has a kernel. We understand well its structure, the kernel  is defined by differences of states $|a\rangle - |a'\rangle$
related by the transformation (\ref{ambi}). The continuum quantum holonomy operators separate some elements of the kernel, 
therefore the operators are not intertwined by the map between the representations. That difference between the two representations  
is caused by non--generic intersections between  loops and triangulations.      

Our continuum representation contains vacua, that is states such that all the quantum holonomy operators eigenvalues are trivial 
(they are the identity).  In addition to the state $|0\rangle$ defined by the zero distributional 1-form $a$ (corresponding also to  $a=d\lambda$, 
for  $\lambda$ constant in each cell of the triangulation) the same property is shared by every state $|a\rangle$  described by  (\ref{vacua})
and  the  discussion laid out below that equality.   The map between our representation and that of DG kills that ambiguity. In particular,  
all the states  that have  trivial holonomy  eigenvalues  are mapped into a single BF vacuum. 

The ambiguity in reconstructing a connection modulo gauge transformations from its parallel transport is a  peculiar property
of the distributional connections we are using. In the case of the states whose eigenvalues  define  non distributional  connections 
there is no ambiguity.  The functions corresponding to $|a\rangle$ and $|a'\rangle$, respectively, are different unless $a$ and
$a'$  differ exactly by  a (distributional) gauge transformation.  The representation derived from the continuum theory is sensitive on that issue, 
while the  DG representation is defined to ignore that subtlety.         
 
The choice of the Abelian gauge group $U(1)$ simplified  our work significantly. In  a non--Abelian group case, the $E$ fields   are not 
gauge invariant any more. In the flux formula, they should be replaced by ones parallel--transported to a single point. 
Such observables are well known in LQG, however building a Hilbert space from them is an unsolved problem. One of the major difficulties is that
the non--Abelian fluxes do not commute.  Still, we anticipate that 
also in that case a manifestly continuum approach defined along the lines  of the current paper is possible.    
\section{Acknowledgements}
This work was supported by the grant of Polish Narodowe Centrum Nauki nr 2011/02/A/ST2/00300.

\end{document}